\documentclass[review]{elsarticle}

\usepackage{lineno,hyperref}

\usepackage{xcolor,listings}
\definecolor{mlgreen}{rgb}{.035,.6,.251}
\definecolor{mlviolett}{rgb}{.643,.259,.804}
\lstdefinestyle{mlab}{language=Matlab, numbers=left, numberstyle=\tiny,
	basicstyle={\ttfamily},%
	keywordstyle={\color{blue}},%
	commentstyle=\color{mlgreen},%
	stringstyle=\color{mlviolett},%
	breaklines=true,
}

\journal{arXiv}

\usepackage{amsmath}

\begin{document}

\begin{frontmatter}

\title{Reciprocal first-order second-moment method}

\author{Benedikt Kriegesmann}
\author{Julian K. L\"udeker}
\address{Hamburg University of Technology}
\address{benedikt.kriegesmann@tuhh.de}

\begin{abstract}
This paper shows a simple parameter substitution, which makes use of the reciprocal relation of typical objective functions with typical random parameters. 
Thereby, the accuracy of first-order probabilistic analysis improves significantly at almost no additional computational cost. 
The parameter substitution requires a transformation of the stochastic distribution of the substituted parameter, which is explained for different cases.
\end{abstract}


\end{frontmatter}


\section{Introduction}

\textsc{Taylor} series approximations of an objective function allow for a very fast determination of the mean and variance of this objective function and are therefore frequently used in robust design optimization. 
While most researchers use second-order approximations (see e.g. \cite{doltsinis_robust_2004,asadpoure_robust_2011}), the authors showed that using a first-order approximation is also suitable for certain cases at much less computational cost \cite{kriegesmann_robust_2019}. 
For second-order approaches, the computational cost at least scales with the number of random parameters, while the first-order approximation presented in \cite{kriegesmann_robust_2019} required solving only two systems of equations.

Typical objective functions are compliance of a structure, the displacement at a certain point, or the maximum stress. 
Typical random parameters are material properties (e.g. \textsc{Young}'s modulus or yield stress), geometric measures (e.g. shape or thickness), and loads (direction and/or magnitude). 
In linear mechanics, displacement and stress are indeed linearly dependent on the load, and the compliance is quadratically dependent on the load magnitude. 
However, for the other random parameters mentioned, this relationship is reciprocal.

The benefits of using reciprocal approximations has already been identified by Schmit and Farshi \cite{schmit_approximation_1974} and is still widely used in the context of structural optimization \cite{fleury_claude_structural_1986,svanberg_method_1987}. 
For probabilistic analyses however, the reciprocal relation has only been utilized in the work of Fuchs and Shabtay \cite{fuchs_reciprocal_2000} (to the best of the authors' knowledge).
They used the reciprocal relation for the perturbation method. Hence, they expanded the equilibrium equation for random stiffness to get the stochastic moments of the displacement vector. 

In this paper, a much simpler formulation is presented to consider the reciprocal relationship in a \textsc{Taylor} expansion of the objective function, which is formulated for arbitrary objective functions and arbitrary random parameters. 
The major implication in using this approach (which also holds for the work of Fuchs and Shabtay \cite{fuchs_reciprocal_2000}) is the need for stochastic moments of the reciprocal parameters. 
Therefore, the paper shows how to determine the requires stochastic moments of the reciprocal parameters for different cases.

\section{First-order second-moment method for mean and variance approximation}
\label{sec:FOSM}

Firstly, the well-known approach to determine mean and variance of an objective function using a \textsc{Taylor} series is recalled.
Consider the objective function $g(\boldsymbol{x})$, which is a function of the random vector $\boldsymbol{X}$. 
The \textsc{Taylor} series approximation if $g$ is expanded at the mean vector ${\boldsymbol{\mu}}_X$ of $\boldsymbol{X}$.
\begin{equation}
	\begin{aligned}
		g\left( {\boldsymbol{x}} \right) & = g\left( {{{\boldsymbol{\mu }}_X}} \right) + \sum\limits_{i = 1}^n {\frac{{\partial g\left( {{{\boldsymbol{\mu }}_X}} \right)}}{{\partial {x_i}}}\left( {{x_i} - {\mu _{{X_i}}}} \right)} \\
		& + \frac{1}{2}\sum\limits_{i = 1}^n {\sum\limits_{j = 1}^n {\frac{{{\partial ^2}g\left( {{{\boldsymbol{\mu }}_X}} \right)}}{{\partial {x_i}\,\partial {x_j}}}\left( {{x_i} - {\mu _{{X_i}}}} \right)\left( {{x_j} - {\mu _{{X_j}}}} \right)} }  +  \ldots
	\end{aligned} 
	\label{eq:Taylor_gx}
\end{equation}
Inserting the first-order terms of (\ref{eq:Taylor_gx}) in the integrals that determine the mean $\mu_g$ and the variance $\sigma_g^2$ yields
\begin{equation}
	\begin{aligned}
		{\mu _g} & = \int\limits_{ - \infty }^\infty  {g\left( {\boldsymbol{x}} \right){f_{\boldsymbol{X}}}\left( {\boldsymbol{x}} \right)d{\boldsymbol{x}}}  \approx g\left( {{{\boldsymbol{\mu }}_X}} \right) \\
		\sigma _g^2 & = \int\limits_{ - \infty }^\infty  {{{\left[ {g\left( {\boldsymbol{x}} \right) - {\mu _g}} \right]}^2}{f_{\boldsymbol{X}}}\left( {\boldsymbol{x}} \right)d{\boldsymbol{x}}}  \approx \sum\limits_{i = 1}^n {\sum\limits_{j = 1}^n {\frac{{\partial g\left( {{{\boldsymbol{\mu }}_X}} \right)}}{{\partial {x_i}}}\frac{{\partial g\left( {{{\boldsymbol{\mu }}_X}} \right)}}{{\partial {x_j}}}{\mathop{\rm cov}} \left( {{X_i},{X_j}} \right)} }
	\end{aligned}
	\label{eq:FOSM_x}
\end{equation}
For a second-order approach, e.g. the second-order fourth-moment (SOFM) method, refer for instance \cite{kriegesmann_fast_2011}.

From eqs. (\ref{eq:FOSM_x}) it becomes obvious that the approximation of $g$ by the \textsc{Taylor} series should be acceptable in areas where $f_{\boldsymbol{X}}({\boldsymbol{x}})$ is significantly different from zero, which is typically the case in the region close to ${\boldsymbol{\mu}}_X$.

\section{Reciprocal first-order approximation by parameter substitution}
\label{sec:reciFOSM}

The basic idea of the reciprocal approach is to substitute the original random vector $\boldsymbol{X}$ by $\boldsymbol{Z}$, where for each $i$-th entry 
\begin{equation}
	{z_i} = \frac{1}{{{x_i}}} \Leftrightarrow {x_i} = \frac{1}{{{z_i}}}
	\label{eq:x_to_z}
\end{equation}
Now, the objective function $g\left( {{\boldsymbol{x}}\left( {\boldsymbol{z}} \right)} \right)$ is expanded in ${\boldsymbol{z}}$ at ${\boldsymbol{\mu}}_Z$.
\begin{equation}
	g\left( {{\boldsymbol{x}}\left( {\boldsymbol{z}} \right)} \right) = g\left( {\boldsymbol{x}}({{{\boldsymbol{\mu }}_Z}}) \right) + \sum\limits_{i = 1}^n {\frac{{\partial g\left( {\boldsymbol{x}}({{{\boldsymbol{\mu }}_Z}}) \right)}}{{\partial {z_i}}}\left( {{z_i} - {\mu _{{Z_i}}}} \right)}  +  \ldots
	\label{eq:Taylor_gz}
\end{equation}
Here, ${\boldsymbol{\mu}}_Z$ is the mean vector of ${\boldsymbol{Z}}$. Inserting (\ref{eq:Taylor_gz}) into eqs. (\ref{eq:FOSM_x}), yields
\begin{equation}
	\begin{aligned}
		{\mu _g} & \approx g\left( {\boldsymbol{x}}({{{\boldsymbol{\mu }}_Z}}) \right) \\
		\sigma _g^2 & \approx \sum\limits_{i = 1}^n {\sum\limits_{j = 1}^n {\frac{{\partial g\left( {\boldsymbol{x}}({{{\boldsymbol{\mu }}_Z}}) \right)}}{{\partial {z_i}}}\frac{{\partial g\left( {\boldsymbol{x}}({{{\boldsymbol{\mu }}_Z}}) \right)}}{{\partial {z_j}}}{\mathop{\rm cov}} \left( {{Z_i},{Z_j}} \right)} }
	\end{aligned}
	\label{eq:FOSM_z}
\end{equation}
The derivatives of (\ref{eq:x_to_z}) equals 
\begin{equation}
	\frac{{\partial {x_i}}}{{\partial {z_j}}} = \left\{ \begin{array}{ccccc}
	- \frac{1}{{z_i^2}}\quad  & \forall i = j\\
	0\quad  & \forall i \ne j
	\end{array} \right.
	\label{eq:dxdz}
\end{equation}
The derivative of $g$ with respect to $z_i$ therefore is
\begin{equation}
	\frac{{\partial g\left( {\boldsymbol{z}} \right)}}{{\partial {z_i}}} = \frac{{\partial g}}{{\partial {x_i}}}\frac{{\partial {x_i}}}{{\partial {z_i}}}
	\label{eq:dgdz}
\end{equation}
What still needs to be determined for evaluating (\ref{eq:FOSM_z}) is the mean vector ${\boldsymbol{\mu}}_Z$ at which $g\left( {{\boldsymbol{x}}\left( {\boldsymbol{z}} \right)} \right)$ and its derivatives are evaluated, and the covariance ${\mathop{\rm cov}} \left( {{Z_i},{Z_j}} \right)$. Determining these stochastic moments based on (\ref{eq:x_to_z}) is less obvious than it may seem, and therefore it is discussed in the next section. 

The authors initially tried to express $\mu_g$ and $\sigma_g^2$ purely in terms of stochastic moments of $\boldsymbol{X}$ by a reciprocal expansion. However, this turned out not to be possible, but also not necessary.

\section{Transformation to reciprocal random parameter}

This section shows how ${\boldsymbol{\mu}}_Z$ and ${\mathop{\rm cov}} \left( {{Z_i},{Z_j}} \right)$ can be determined firstly, in case the distribution of $\boldsymbol{X}$ is given explicitly and secondly, in case measurement data of $\boldsymbol{X}$ are available. 

\subsection{Case 1 – distribution given}
\label{sec:reciRand_distGiven}
Given the probability density function distribution $f_{X}$ of a single random parameter $X$, the probability density function of $Z$ (which is related by (\ref{eq:x_to_z})) is given by
\begin{equation}
	{f_Z}\left( z \right) = \frac{1}{{{z^2}}}{f_X}\left( {\frac{1}{z}} \right)
	\label{eq:pdf_trans}
\end{equation}
Based on (\ref{eq:pdf_trans}), the mean and the variance of $Z$ can be determined. 

what is however required for applying the FOSM approach is the mean and the variance of $Z$, which are given by 
\begin{equation}
	{\mu _Z} = \int\limits_0^\infty  {\frac{1}{z}{f_X}\left( {\frac{1}{z}} \right)\;dz}
	\label{eq:mu_Z}
\end{equation}
and
\begin{equation}
	\sigma _Z^2 = \int\limits_0^\infty  {{f_X}\left( {\frac{1}{z}} \right)\;dz}  - \mu _Z^2
	\label{eq:sigma2_Z}
\end{equation}
The Appendix provides the few lines of Matlab code which are required to solve (\ref{eq:mu_Z}) and (\ref{eq:sigma2_Z}).
However, solving these integrals is not always easy and sometimes even impossible (e.g. for the standard \textsc{Gauss} distribution).
On the other hand, for several distributions $f_X$ the distribution of the reciprocal variable $f_Z$ is known, e.g. for 
uniform distribution, \textsc{Cauchy} distribution, $F$-distribution, gamma distribution \cite{forbes_statistical_2011}.
This is demonstrated for the very simple case of $F$-distribution, which has the two parameters $m$ and $n$. Mean and variance of the $F$-distribution are given by
\begin{equation}
	{\mu _{F{\rm{ - dist}}}} = \frac{n}{{n - 2}} \quad \quad \quad \quad \sigma _{F{\rm{ - dist}}}^2 = \frac{{2{n^2}\left( {m + n - 2} \right)}}{{m{{\left( {n - 2} \right)}^2}\left( {n - 4} \right)}}
	\label{eq:mean_var_Fdist}
\end{equation}
If $X$ follows $F$-distribution, i.e. $X \sim F(m,n)$, then $Z \sim F(n,m)$.

This section only looked at single random parameters. If the entries of the random vector $\boldsymbol{X}$ are independent, the transformation of the distribution can be done component-wise. However, if the entries of $\boldsymbol{X}$ are dependent, their covariance has to be considered. In case $\boldsymbol{X}$ is \textsc{Gaussian}, it can be transformed easily to independent parameters. In general however, this may not be possible and makes it difficult to determine $f_{\boldsymbol{Z}}$ analytically. A workaround in such case is to generate random realizations $\boldsymbol{x}$ of $\boldsymbol{X}$ and use the approach given in the following section.

\subsection{Case 2 – data give}

Given $n$ measurement data (or realizations) ${\boldsymbol{x}^{(k)}}$ of the random vector $\boldsymbol{X}$, each $i$-th entry of each $k$-th realization can be transformed to ${z_i^{(k)}} = \frac{1}{{{x_i^{(k)}}}}$.

Then, mean vector and covariance matrix of ${\boldsymbol{Z}}$ are determined by the well-known empirical estimators
\begin{equation}
	\begin{aligned}
		{\boldsymbol{\mu}}_Z & \approx \frac{1}{n} \sum_{k=1}^{n} {\boldsymbol{z}^{(k)}} \\
		{\mathop{\rm cov}} \left( {{Z_i},{Z_j}} \right) & \approx \frac{1}{n-1} \sum_{k=1}^{n} {z_i^{(k)}}
		\left( {{z_i^{(k)}} - {\mu _{{Z_i}}}} \right)\left( {{z_j^{(k)}} - {\mu _{{Z_j}}}} \right)
	\end{aligned} 
\end{equation}

\section{Examples}
\begin{figure}[h]
	\centering
	\begin{minipage}{.4\textwidth}
		\centering
		\includegraphics[width=\textwidth]{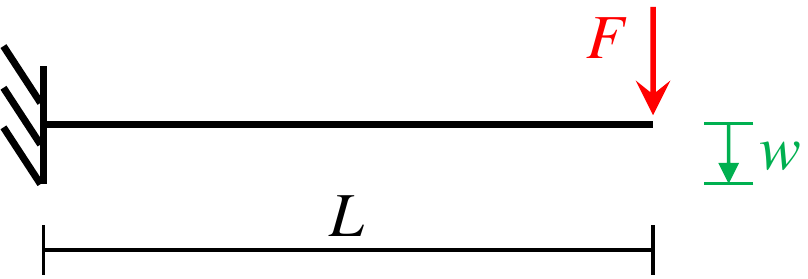}
	\end{minipage}%
	\caption{Cantilever beam example}
	\label{fig:cantilever}
\end{figure}
For demonstrating the effect of using the reciprocal FOSM method given in the previous section, consider the cantilever beam example shown in Figure~\ref{fig:cantilever} with the load $F = 0.1kN$, the length $L = 1000mm$, the \textsc{Young}'s modulus $E = 70kN/mm^2$ and a rectangular cross section with the height $h=30mm$ and width $b=30mm$.
The objective function is the displacement at the tip $w = 4FL^3/(Eh^3b)$.

\begin{figure}[h]
	\centering
	\begin{minipage}{.49\textwidth}
		\centering
		\includegraphics[width=\textwidth]{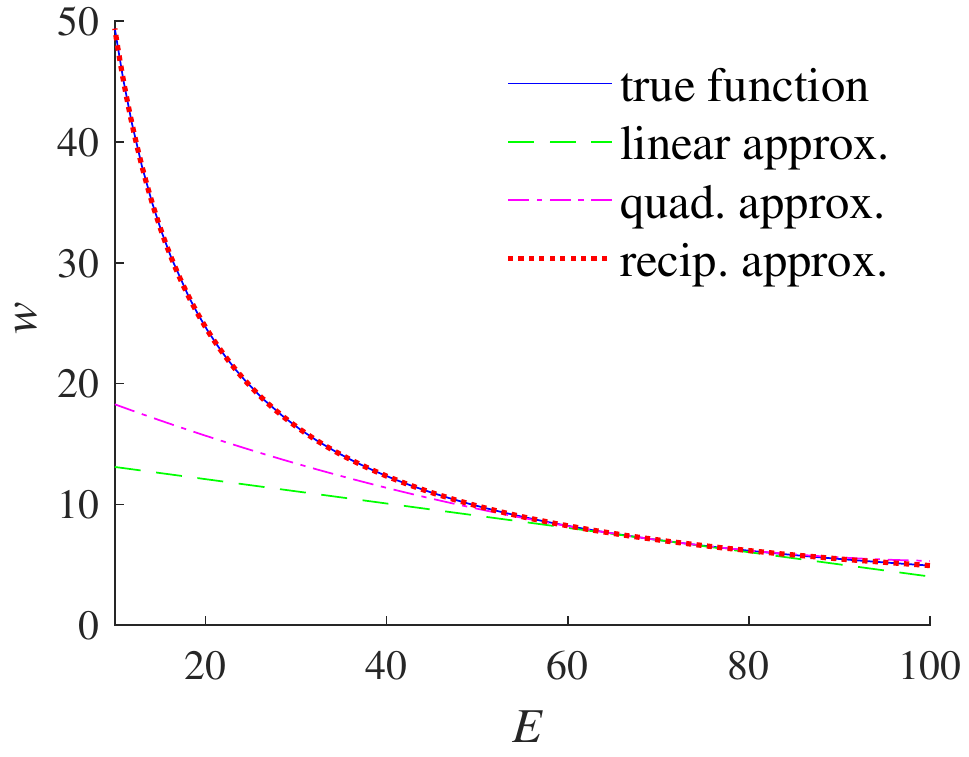}
	\end{minipage}%
	\hspace{0.01\textwidth}
	\begin{minipage}{.49\textwidth}
		\centering
		\includegraphics[width=\textwidth]{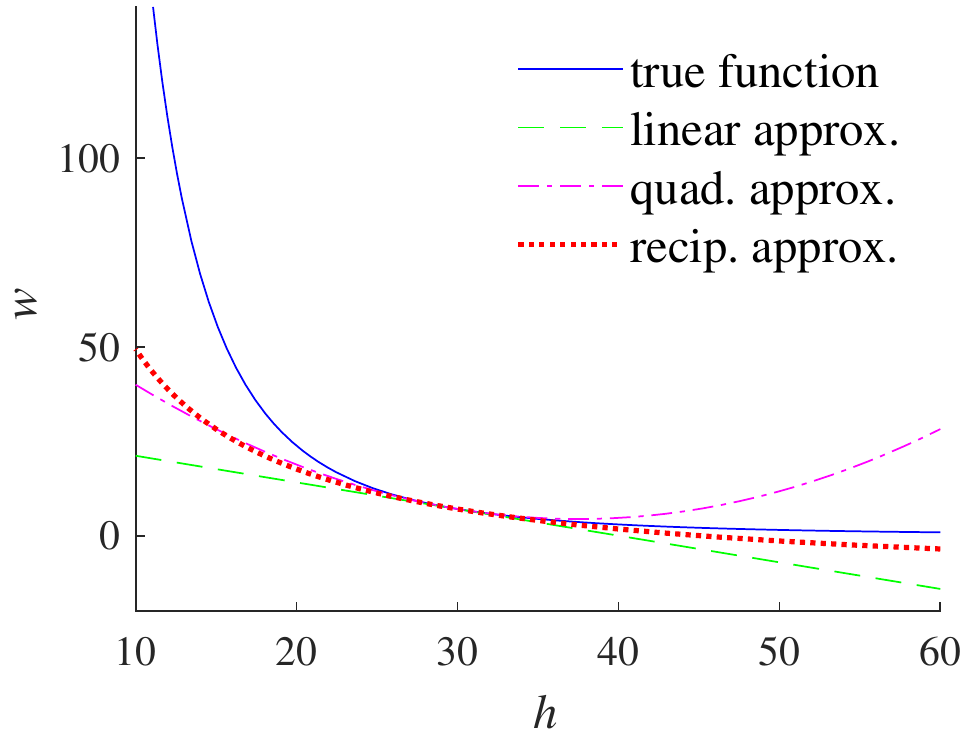}
	\end{minipage}
	\caption{Displacement at tip $w$ as a function of \textsc{Young}'s modulus (left) and height (right), showing the first-order, second-order and first-order reciprocal approximation}
	\label{fig:w_over_Eh}
\end{figure}

In the following, the \textsc{Young}'s modulus $E$ and the height $h$ will be considered as random. 
As depicted in Figure~\ref{fig:w_over_Eh}, for this simple example the objective function $w$ is indeed inversely proportional with respect to $E$, where the dependency of $w$ and $h$ is of higher-order nonlinearity.

\subsection{Case 1 – distribution given}

The \textsc{Young}'s modulus is now expressed as $E = \alpha \cdot E_0$, where $E_0$ is the nominal value and $\alpha$ is a random parameter that follows $F$-distribution with $m=25$ and $n=100$. Using eqs. (\ref{eq:mean_var_Fdist}), the mean value and standard deviation of $E$ the are determined to equal $\mu_E = 71.4kN/mm^2$ and $\sigma_E = 22.9kN/mm^2$. In this example, the coefficient of variation equals $CoV = 0.32$, which is unrealistic large for most materials and has been chose for demonstration purpose.
As stated in section~\ref{sec:reciRand_distGiven}, the reciprocal random parameter $Z=1/\alpha$ follows $F$-distribution with $m=100$ and $n=25$, 

Table~\ref{tab:resCase1} summarizes the results of applying the FOSM method, the SOFM method, the reciprocal FOSM (recFOSM) method introduced in section~\ref{sec:reciFOSM} and the Monte Carlo method with $10^5$ samples.
Since the deflection $w$ is indeed reciprocal with respect to the \textsc{Young}'s modulus $E$, the reciprocal FOSM approach provides almost exactly the same results as the Monte Carlo simulation at the same computational cost as the standard FOSM method.
\begin{table}[h]
	\centering
	\begin{tabular}{lcccc}
		\hline \hline
		Approach & FOSM & SOFM & recFOSM & Monte Carlo \\
		\hline
		$\mu_w$ & 6.91 & 7.62 & 7.67 & 7.68 \\
		$\sigma_w$ & 2.21 & 2.51 & 2.62 & 2.63 \\
		\hline \hline
	\end{tabular}
	\caption{Mean and standard deviation of the beam displacement $w$ due to a $F$-distributed random \textsc{Young}'s modulus determined by different approaches}
	\label{tab:resCase1}
\end{table}
%

\subsection{Case 2 – data given}

In order to simulate the case that measurement data are given, random realizations are generated using the \textsc{Weibull} distribution. Here, the mean value always equals the nominal value and the variance is determined from a prescribed coefficient of variation, which varies throughout the section.

\begin{figure}[h]
	\centering
	\begin{minipage}{.49\textwidth}
		\centering
		\includegraphics[width=\textwidth]{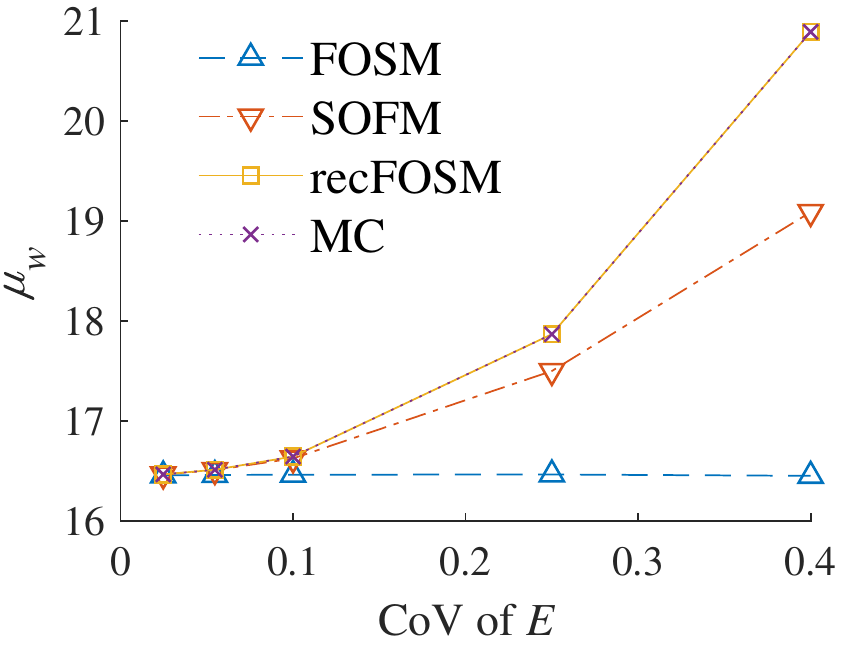}
	\end{minipage}%
	\hspace{0.01\textwidth}
	\begin{minipage}{.49\textwidth}
		\centering
		\includegraphics[width=\textwidth]{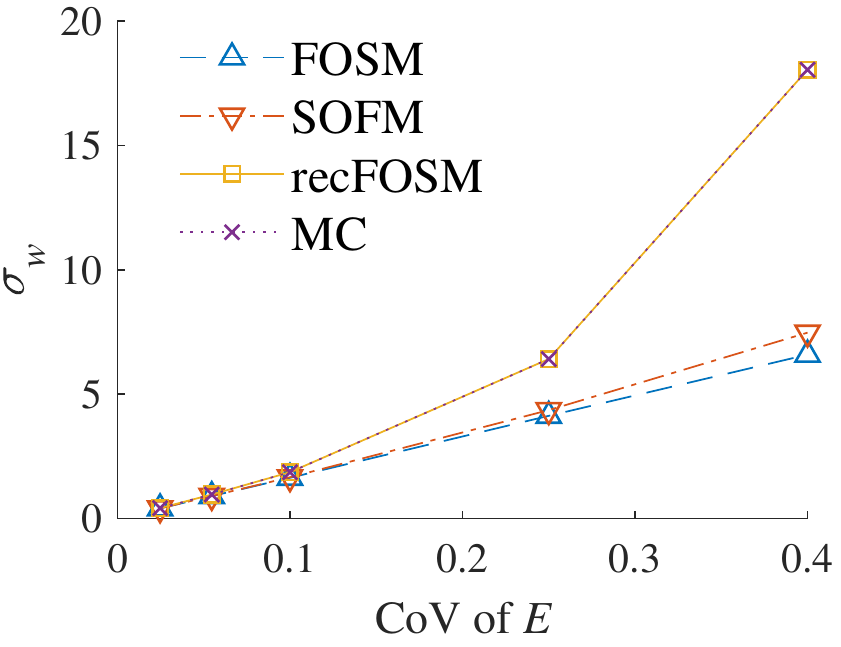}
	\end{minipage}
	\caption{Mean $\mu_w$ (left) and standard deviation $\sigma_w$ (right) of the beam displacement due to a random \textsc{Young}'s modulus for increasing coefficient of variation CoV}
	\label{fig:randE_over_CoV}
\end{figure}

Firstly, consider the \textsc{Young}'s modulus as random parameter. The coefficient of variation takes the values of $\text{CoV}  \in [0.025, 0.05, 0.1, 0.25, 0.4]$.
Figure~\ref{fig:randE_over_CoV} shows the results of applying the four approaches considered for different coefficients of variation of the \textsc{Young}'s modulus. Table~\ref{tab:resCase2_E} gives numerical values for the most extreme cases. The results show that for a small CoVs, FOSM and SOFM approaches are accurate, while for large CoVs they deviate strongly from the Monte Carlo solution (using $10^5$ samples). The reciprocal FOSM result however is always accurate.
\begin{table}[h]
	\centering
	\begin{tabular}{lcccc}
		\hline \hline
		Approach & FOSM & SOFM & recFOSM & Monte Carlo \\
		\hline
		$\mu_w$ for CoV$=0.025$ & 16.46 & 16.47 & 16.47 & 16.47 \\
		$\sigma_w$ for CoV$=0.025$& 0.407 & 0.407 & 0.418 & 0.418 \\
		\hline
		$\mu_w$ for CoV$=0.4$ & 16.46 & 17.49 & 17.85 & 17.85 \\
		$\sigma_w$ for CoV$=0.4$ & 4.11 & 4.34 & 6.37 & 6.37 \\
		\hline \hline
	\end{tabular}
	\caption{Mean and standard deviation of the beam displacement $w$ due to a random \textsc{Young}'s modulus determined by different approaches and different coefficients of variation}
	\label{tab:resCase2_E}
\end{table}
%
%
%
%
%
%

\begin{figure}[h]
	\centering
	\begin{minipage}{.49\textwidth}
		\centering
		\includegraphics[width=\textwidth]{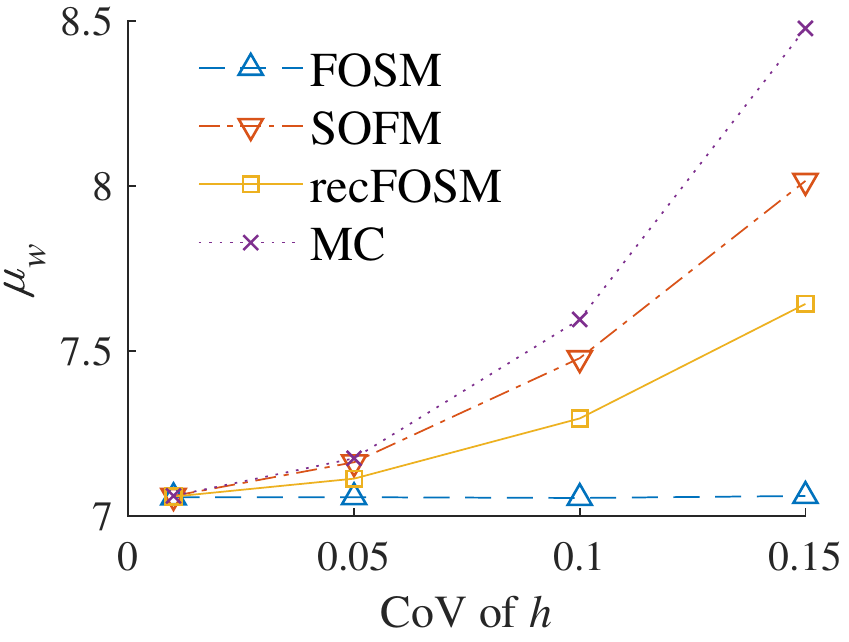}
	\end{minipage}%
	\hspace{0.01\textwidth}
	\begin{minipage}{.49\textwidth}
		\centering
		\includegraphics[width=\textwidth]{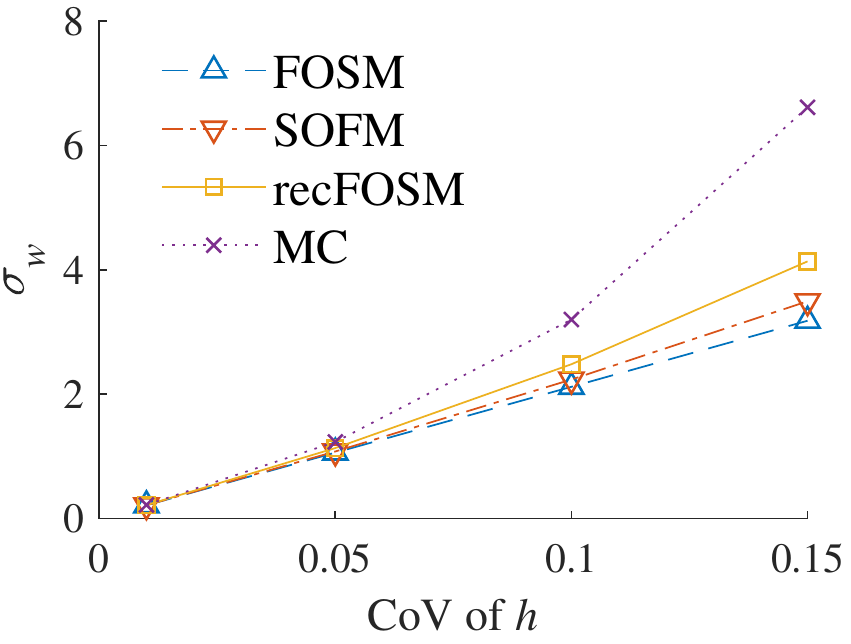}
	\end{minipage}
	\caption{Mean $\mu_w$ (left) and standard deviation $\sigma_w$ (right) of the beam displacement due to a random cross section height for increasing coefficient of variation CoV}
	\label{fig:randH_over_CoV}
\end{figure}

Next, the cross section height $h$ is considered as random parameter. Its coefficient of variation is chosen to take values of $\text{CoV}  \in [0.01, 0.05, 0.1, 0.15]$. 
The results are given in Table~\ref{tab:example_results} and shown in Figure~\ref{fig:randH_over_CoV}.
Due to the strong nonlinear influence of $h$ on the deflection $w$, the results of FOSM, SOFM and reciprocal FOSM deviate from the Monte Carlo solution already for small CoVs. While SOFM provides better results for the mean $\mu_w$, the reciprocal FOSM is more accurate for the standard deviation $\sigma_w$. 
\begin{table}[h]
	\centering
	\begin{tabular}{lcccc}
		\hline \hline
		Approach & FOSM & SOFM & recFOSM & Monte Carlo \\
		\hline
		$\mu_w$ for CoV$=0.01$ & 7.056 & 7.061 & 7.059 & 7.061 \\
		$\sigma_w$ for CoV$=0.01$ & 0.213 & 0.212 & 0.215 & 0.218 \\
		\hline
		$\mu_w$ for CoV$=0.15$ & 7.06 & 8.02 & 7.64 & 8.48 \\
		$\sigma_w$ for CoV$=0.15$ & 3.18 & 3.49 & 4.13 & 6.61 \\
		\hline \hline
	\end{tabular}
	\caption{Mean and standard deviation of the beam displacement $w$ due to a random cross section height determined by different approaches and different coefficients of variation}
	\label{tab:example_results}
\end{table}
%
%
%
%
%

\section{Conclusion}

The presented reciprocal FOSM method requires the same number of function evaluations and the same derivatives as the standard FOSM method. 
Hence, the computational cost is the same and therefore much less than for a second-order approach or Monte Carlo simulations.
For parameters which have a reciprocal relationship to the objective function, the reciprocal FOSM method provides exact results. 
If the objective function is non-linear with respect to the considered parameter, the reciprocal FOSM method still provides much more accurate results than the FOSM approach and can even be more accurate than a second-order approach.
Especially for the use in robust design optimization, the reciprocal FOSM method is a promising approach as it accurately determines the variance of an objective at very low computational costs.

\section*{Appendix}
This Appendix provides the Matlab code for determining the mean and variance of the the reciprocal random variable $Z$ from a given probability density function of $X$. The example is given for a Weibull distribution. The inverse cumulative distribution function (CDF) is used for the validation via Monte Carlo sampling. The most challenging part of applying the example to other distributions is the generation of samples for the validation. Since for many distributions the inverse CDF is unknown, it has to be determined numerically or an acceptance rejection method has to be used. On request, the author can provide further examples, such as for log-normal, normal or $\beta$-distribution.
\begin{lstlisting}[style=mlab] 
% Weibull distribution
a=3; b=5;
pdfX = @(x) a*b* x.^(b-1) .* exp(-a* x.^b );
invCDFx = @(u) ( -1/a * log(1-u) ).^(1/b);
%
% solve integral for mean
fun = @(z) 1./z.*pdfX(1./z);
mu_Z = integral(fun,0,Inf)
% ... and variance
fun = @(z) pdfX(1./z);
sigma2_Z = integral(fun,0,Inf) - mu_Z^2
%
% validation with Monte Carlo
nos = 1000000; % number of samples
xr = invCDFx(rand(nos,1));
zr = 1./xr;
mu_Z_MC = mean(zr)
sigma2_Z_MC = var(zr)
\end{lstlisting}


\bibliography{EBHF}

\begin{thebibliography}{1}
\expandafter\ifx\csname url\endcsname\relax
  \def\url#1{\texttt{#1}}\fi
\expandafter\ifx\csname urlprefix\endcsname\relax\def\urlprefix{URL }\fi
\expandafter\ifx\csname href\endcsname\relax
  \def\href#1#2{#2} \def\path#1{#1}\fi

\bibitem{doltsinis_robust_2004}
I.~Doltsinis, Z.~Kang,
  \href{http://www.sciencedirect.com/science/article/pii/S0045782504000787}{Robust
  design of structures using optimization methods}, Computer Methods in Applied
  Mechanics and Engineering 193~(23) (2004) 2221--2237.
\newblock \href {http://dx.doi.org/10.1016/j.cma.2003.12.055}
  {\path{doi:10.1016/j.cma.2003.12.055}}.
\newline\urlprefix\url{http://www.sciencedirect.com/science/article/pii/S0045782504000787}

\bibitem{asadpoure_robust_2011}
A.~Asadpoure, M.~Tootkaboni, J.~K. Guest,
  \href{http://www.sciencedirect.com/science/article/pii/S004579491000266X}{Robust
  topology optimization of structures with uncertainties in stiffness –
  {Application} to truss structures}, Computers \& Structures 89~(11) (2011)
  1131--1141.
\newblock \href {http://dx.doi.org/10.1016/j.compstruc.2010.11.004}
  {\path{doi:10.1016/j.compstruc.2010.11.004}}.
\newline\urlprefix\url{http://www.sciencedirect.com/science/article/pii/S004579491000266X}

\bibitem{kriegesmann_robust_2019}
B.~Kriegesmann, J.~K. Lüdeker,
  \href{https://doi.org/10.1007/s00158-019-02216-8}{Robust compliance topology
  optimization using the first-order second-moment method}, Structural and
  Multidisciplinary Optimization 60~(1) (2019) 269--286.
\newblock \href {http://dx.doi.org/10.1007/s00158-019-02216-8}
  {\path{doi:10.1007/s00158-019-02216-8}}.
\newline\urlprefix\url{https://doi.org/10.1007/s00158-019-02216-8}

\bibitem{schmit_approximation_1974}
L.~A. SCHMIT, B.~FARSHI, \href{https://doi.org/10.2514/3.49321}{Some
  {Approximation} {Concepts} for {Structural} {Synthesis}}, AIAA Journal 12~(5)
  (1974) 692--699, publisher: American Institute of Aeronautics and
  Astronautics \_eprint: https://doi.org/10.2514/3.49321.
\newblock \href {http://dx.doi.org/10.2514/3.49321}
  {\path{doi:10.2514/3.49321}}.
\newline\urlprefix\url{https://doi.org/10.2514/3.49321}

\bibitem{fleury_claude_structural_1986}
{Fleury Claude}, {Braibant Vincent},
  \href{https://onlinelibrary.wiley.com/doi/abs/10.1002/nme.1620230307}{Structural
  optimization: {A} new dual method using mixed variables}, International
  Journal for Numerical Methods in Engineering 23~(3) (1986) 409--428.
\newblock \href {http://dx.doi.org/10.1002/nme.1620230307}
  {\path{doi:10.1002/nme.1620230307}}.
\newline\urlprefix\url{https://onlinelibrary.wiley.com/doi/abs/10.1002/nme.1620230307}

\bibitem{svanberg_method_1987}
K.~Svanberg,
  \href{http://onlinelibrary.wiley.com/doi/10.1002/nme.1620240207/abstract}{The
  method of moving asymptotes - a new method for structural optimization},
  International Journal for Numerical Methods in Engineering 24~(2) (1987)
  359--373.
\newblock \href {http://dx.doi.org/10.1002/nme.1620240207}
  {\path{doi:10.1002/nme.1620240207}}.
\newline\urlprefix\url{http://onlinelibrary.wiley.com/doi/10.1002/nme.1620240207/abstract}

\bibitem{fuchs_reciprocal_2000}
M.~B. Fuchs, E.~Shabtay,
  \href{http://www.sciencedirect.com/science/article/pii/S0960077998002422}{The
  reciprocal approximation in stochastic analysis of structures}, Chaos,
  Solitons \& Fractals 11~(6) (2000) 889--900.
\newblock \href {http://dx.doi.org/10.1016/S0960-0779(98)00242-2}
  {\path{doi:10.1016/S0960-0779(98)00242-2}}.
\newline\urlprefix\url{http://www.sciencedirect.com/science/article/pii/S0960077998002422}

\bibitem{kriegesmann_fast_2011}
B.~Kriegesmann, R.~Rolfes, C.~Hühne, A.~Kling, Fast {Probabilistic} {Design}
  {Procedure} for {Axially} {Compressed} {Composite} {Cylinders}, Composites
  Structures 93 (2011) 3140--3149.
\newblock \href {http://dx.doi.org/10.1016/j.compstruct.2011.06.017}
  {\path{doi:10.1016/j.compstruct.2011.06.017}}.

\bibitem{forbes_statistical_2011}
C.~Forbes, M.~Evans, N.~Hastings, B.~Peacock,
  \href{https://onlinelibrary.wiley.com/doi/10.1002/9780470627242}{Statistical
  distributions}, 4th Edition, John Wiley \& Sons, 2011.
\newline\urlprefix\url{https://onlinelibrary.wiley.com/doi/10.1002/9780470627242}

\end{thebibliography}

\end{document}